\documentclass[12pt]{iopart}

\usepackage{graphicx} 
\usepackage{dcolumn} 
\usepackage{bm} 

\newcommand{\gtequiv}{\lower2pt\hbox{$\:\stackrel{>}{
            \scriptstyle\sim}\:$}}
\newcommand{\ltequiv}{\lower2pt\hbox{$\:\stackrel{<}{
            \scriptstyle\sim}\:$}}

\begin{document}

\title{Superconductivity enhanced conductance fluctuations in few layer graphene nano\-ribbons}

\author{J. Trbovic, N. Minder, F. Freitag, C. Sch\"onenberger}

\address{Department of Physics, University of Basel, Klingel\-bergstrasse 82, 4056 Basel, Switzerland}
\ead{jelena.trbovic@unibas.ch}

\date{\today}

\begin{abstract}
We investigate the mesoscopic disorder induced rms conductance variance $\delta G$ in a few layer
graphene nano\-ribbon (FGNR) contacted by two super\-conducting (S) Ti/Al contacts.
By sweeping the back-gate voltage, we observe pronounced conductance fluctuations superimposed on a linear
background of the two terminal conductance $G$. The linear gate-voltage induced response can be modeled by a set
of inter\-layer and intra\-layer capacitances. $\delta G$ depends on temperature $T$ and source-drain voltage $V_{sd}$.
$\delta G$ increases with decreasing $T$ and $|V_{sd}|$. When lowering $|V_{sd}|$, a pronounced cross-over at a voltage
corresponding to the super\-conducting energy gap $\Delta$ is observed. For $|V_{sd}|\ltequiv \Delta$ the fluctuations
are markedly enhanced. Expressed in the conductance variance $G_{GS}$ of one graphene-super\-conductor (G-S) interface,
values of $0.58\,e^2/h$ are obtained at the base temperature of \mbox{$230$\,mK}.
The conductance variance in the sub-gap region are larger by up to a factor of $1.4-1.8$
compared to the normal state. The observed strong enhancement is due to phase coherent charge transfer
caused by Andreev reflection at the nano\-ribbon-superconductor interface.
\end{abstract}


\maketitle

\section{Introduction}

Graphene nano\-structures~\cite{NovoselovSc04, TheRise} provide the unique opportunity to study
fundamentally new quantum coherent phenomena and, on the other hand, possess an immense potential for
applications~\cite{Gprospects}. The new quantum phenomena in graphene originate from the linear
band structure of the relativistic-like quasi\-particles and their chirality. The demonstration of a tunable
super\-current posed fundamentally new questions regarding the sources of apparent dephasing in this material
since the amplitude of the critical current is lower than expected~\cite{Hee07,XuDu2008}.
Previous studies in disordered graphene devices have shown~\cite{MorozovPRL06, HBHeerscheEPJSpec2007,OjedaPRB2009}
that interference effects such as the universal conductance fluctuation (UCF) and weak localisation are dominant corrections
to the conductance at low temperatures. First studies of UCF and weak localisation were done in the 80's in metallic systems
and 2D semiconducting hetero\-structures~\cite{Lee87}. Conductance fluctuations arising in these systems are universal,
independent of sample size and degree of disorder, reaching values of the order $e^{2}/h$~\cite{Lee87}.
The interest in studying conductance fluctuations in graphene stems from predictions that they cease to be universal
in the coherent state of disordered graphene~\cite{RycerzEPL2007}. In graphene samples a range of sources of disorder
has been identified: the close proximity with the substrate~\cite{FuhrerNNanTech08},
interaction with the leads~\cite{PBlakeContacts, contactLeeNNT08}, rippling of the graphene layer~\cite{GuineaPRB08},
and unintentional doping. The disorder reduces the mobility of the quasi\-particles.
In the strong disorder regime, the variance of the conductance $\delta G$ coincides with the predicted value for
disordered metals, whereas in the weak disorder regime, $\delta G$ is larger than the universal value
due to the absence of back-scattering, characteristic of the honeycomb lattice of graphene~\cite{RycerzEPL2007}.
Additional  information on UCF in the system can be gained by attaching super\-conducting contacts to the
disordered region~\cite{BeenakkerNato}. At the interface between a normal metal $N$ and a super\-conducting electrode $S$
phase sensitive Andreev reflections occur for energies lower than the super\-conducting energy gap $\Delta$.
At the N-S interface, an electron coming from $N$ couples with its time-reversed counterpart to form a Cooper pair which can enter into $S$.
The phase coherent Andreev states at the N-S interface can be destroyed by applying a magnetic field which breaks
time reversal symmetry, by increasing the temperature, or by applying a large source-drain voltage.

In this work we study the two terminal conductance of a few layer graphene nanoribbon (FLGNR)
contacted by Ti/Al leads as a function of back-gate voltage $V_{g}$ and source drain voltage $V_{sd}$.
We find that the conductance variance increases with lowering temperature reaching an amplitude of the order of $e^2/h$ at
zero bias and base temperature of \mbox{$230$\,mK}. By applying a source drain voltage larger than $\Delta /e$
the conductance variance decrease by up to a factor of $1.8$. A characteristic cross-over at an energy corresponding to
$\Delta$ confirms that the observed enhancement is due to Andreev reflection at the graphene-S (G-S) interface.
This finding complements existing work~\cite{HBHeerscheEPJSpec2007,XuDu2008,OjedaPRB2009} by focusing at the bias
dependence of conductance variance.

\section{Sample fabrication and characterisation}

Graphene flakes are prepared by mechanical exfoliation of natural graphite (NGS GmbH, Leinburg, Germany) using
a surface protection tape (SPV 224P from Nitto Denko), followed by the transfer of flakes onto a piece
of highly p-doped Si wafer with a top thermal oxide layer of thickness \mbox{$t=304$\,nm}.
The high-doping of the substrate ensure the possibility to gate the flakes by applying a back-gate voltage $V_{g}$ to the substrate.
After flake transfer the samples are rinsed in solvents to remove the glue residue from the flakes and substrate.
Suitable flakes are selected and localized by an optical microscope with respect to a grid of markers.
The devices were patterned with e-beam lithography and subsequently metallized in an UHV e-beam evaporation system at a
pressure of \mbox{$10^{-7}$\,mbar}, followed by lift-off in acetone. We focused on single and few-layer graphene devices
contacted with super\-conducting aluminium (Al). In particular, we studied the sample shown in the inset of figure~1.
This narrow multi-layer flake was contacted with a Ti/Al/Ti tri-layer ($5$/$40$/$20$\,nm). The purpose of the bottom Ti layer
is to ensure high contact transparency between Al and graphene, whereas the top Ti layer caps the Al underneath. From the SEM image,
we estimated the width $W$ of the flake to be between \mbox{$150$\,nm} and \mbox{$200$\,nm} and the edge-to-edge distance $L$
between the leads to be \mbox{$\sim 225$\,nm}.

\begin{figure}[htb]
  \begin{center}
  \includegraphics[width=0.75\linewidth]{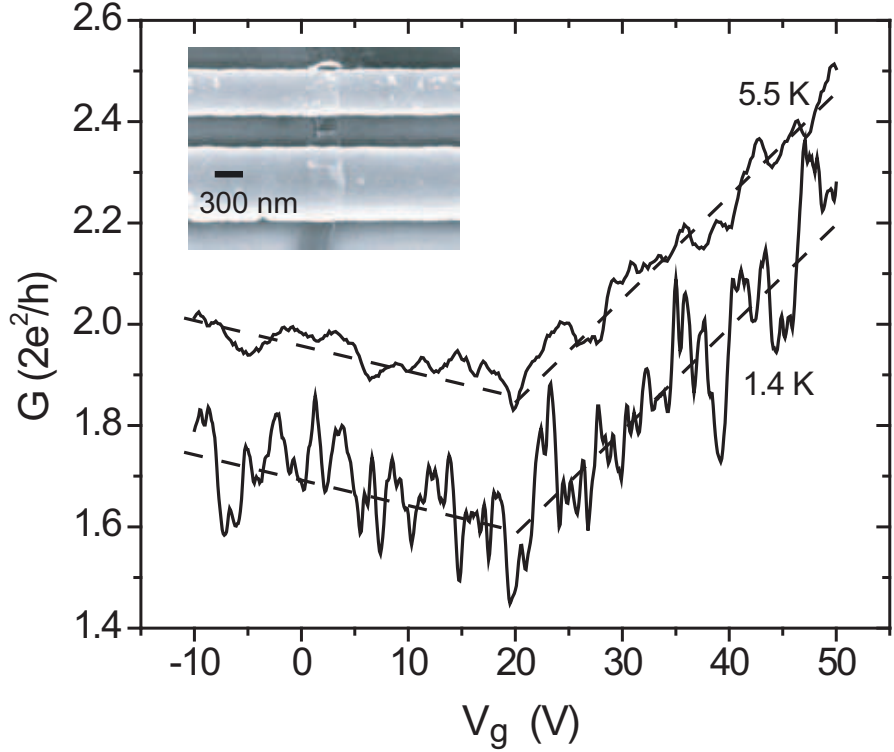}
    \caption{Two terminal linear conductance $G$ taken at \mbox{$5.5$\,K} and \mbox{$1.4$\,K} as a function of the
    back-gate voltage $V_{g}$. The \mbox{$5.5$\,K} curve is shifted by $0.2 \cdot 2e^2/h$ to clearly show the increase
    of the conductance fluctuations upon lowering the temperature.
    The inset shows a scanning electron micrograph of the sample with Ti/Al/Ti electrodes across a few-layer graphene nano\-ribbon.}
  \end{center}
\end{figure}

The back-gate dependence of the two terminal linear differential conductance $G= dI/dV$ taken at \mbox{$1.4$\,K} and \mbox{$5.5$\,K}
(shifted by $0.2 \cdot 2e^{2}/h$ for clarity) in the range of $V_{g}\in [-10V, 50V]$ is shown in figure~1.
Both traces are taken at temperatures where the contact leads are in the normal state. They show a minimum of the conductance
$G_{min}\approx 3.2\,e^2/h$ at \mbox{$20$\,V}, marking the position of the charge neutrality point (CNP).
A small back-gate shift between the two curves of \mbox{$\approx 2$\,V} is observed after cooling the sample from \mbox{$5.5$\,K} to
\mbox{$1.4$\,K}. At \mbox{$5.5$\,K} the conductance linearly depends on $V_{g}$ away from the CNP and shows small aperiodic,
but reproducible fluctuations. The amplitude of the fluctuations becomes larger after lowering the temperature to \mbox{$1.4$\,K}.

The magnitude of the conductance slope $\Delta G/\Delta V_{g}$ (marked with dashed lines in figure~1) can be used to
determine the number of layers as demonstrated in the work of Zhang {\it et al.}~\cite{Zhang05}.
Because of the strong electrostatic screening in the vertical direction of the FLGNR stack, the outermost carbon
layer is affected most by the back-gate voltage. In the inner layers, in contrast, the electric gate-field is strongly
suppressed so that these layers add a nearly gate-independent conductance value proportional to the number of layers to
the total conductance $G$. If one assumes that the electron diffusion constant is the same in all layers,
an estimate for the number of layers $N_L$ in the FLGNR can be obtained from the change in conductance $\Delta G/\Delta V_{g}$
normalized to the minimum conductance $G_{min}$~\cite{Zhang05,TIhn2008}.

\begin{figure}[htb]
  \begin{center}
  \includegraphics[width=0.8\linewidth]{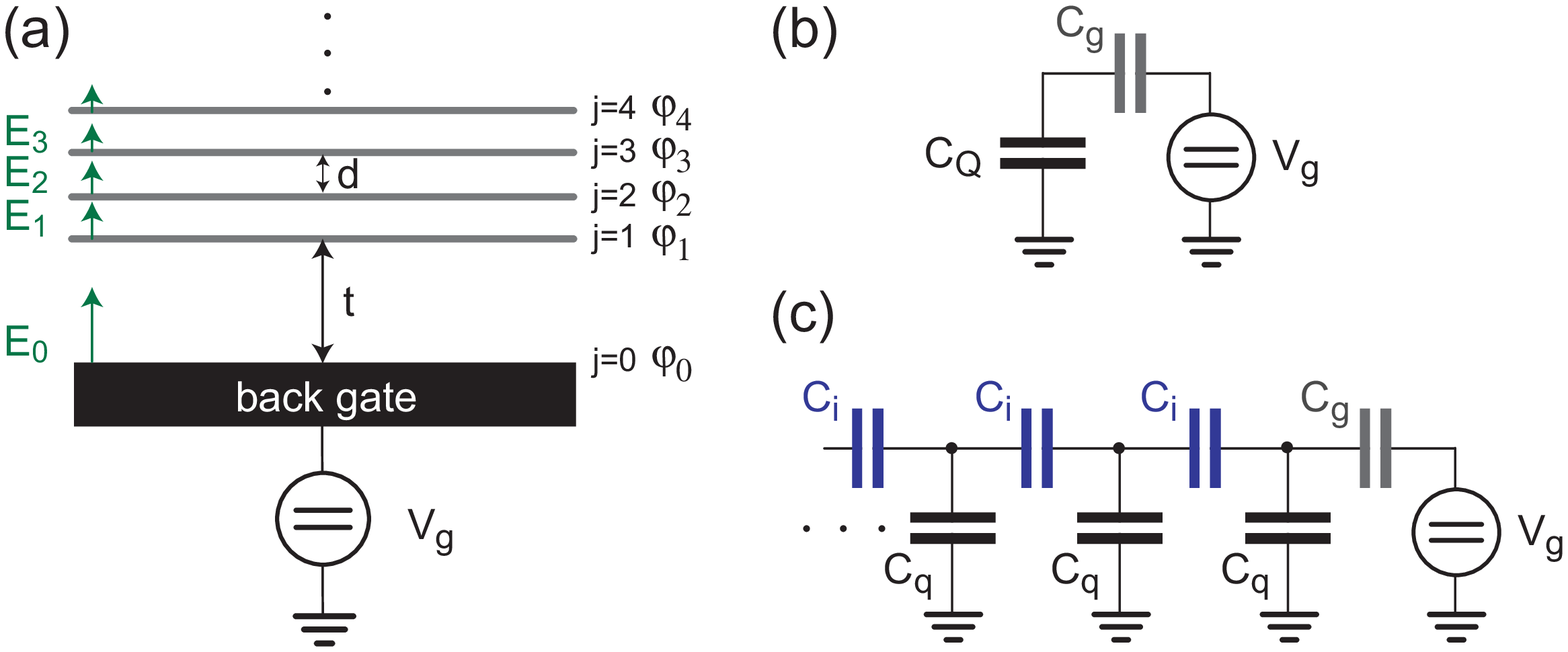}
    \caption{(a) Schematics of the model to calculate the electric-field screening in stack of few layer graphene which
    is gated by an electro\-static back-gate voltage $V_g$. $E_j$ and $\phi_j$ denote the electric field and the
    electro\-static potential in layer $j$.
    (b) The circuit can be reduced to a series connection of two capacitances, where $C_Q$ is an effective quantum capacitance
    and $C_g$ the gate capacitance.
    (b) The full circuit diagram where $C_g$, $C_i$, and $C_q$ denote the gate, graphite inter\-layer, and
    graphene intra\-layer capacitance, respectively. The latter capacitance is sometimes also termed quantum capacitance.
    }
  \end{center}
\end{figure}

To be more explicit, we use the two-dimensional model described in figure~2a. The back-gate is spaced by a
distance $t$ from an infinite set of graphene layers $j=1,2,3,\dots$. The inter\-layer thickness
between the graphene layers is taken to be $d$. $E_j$, $Q_j$, and $\phi_j$ denote the electrical field,
the areal density of the excess charge measured from the CNP, and the electro\-static potential in the different graphene
layers ($j\geq 1$) or back-gate (j=0), respectively. For simplicity we set the dielectric constant to the vacuum
permittivity $\epsilon_0$. This assumption can be relaxed afterwards by replacing the relevant parameter by the
correct back-gate capacitance $C_g$.

The difference in the electric fields is determined by the excess charge according to
\begin{equation}
  Q_j=\epsilon_0 (E_j-E_{j-1}) \mathrm{ .}
  \label{EQ_1}
\end{equation}
We assume that the graphene stack has a back contact (in our case realized by the source and drain contacts),
which are set to zero, whereas the back-gate is biased to an electro\-chemical potential $V_g$.
Because electrons can be exchanged between the different graphene layers by tunnelling, the electro\-chemical potentials
of all layers are equal and zero in the thermo\-dynamic limit. As a consequence the chemical potential $\mu_j$ of layer
$j$ is given by the negative of the electrostatic potential $\phi_j$, i.e. $\mu_j=-e\phi_j$ with $e$ the unit of charge.
On the other hand, the chemical potential $\mu_j$ is determined by the excess charge density according to
$Q_j/e N_j$, where the new symbol $N_j$ denotes the areal density of states in graphene layer $j$. Taken together, we
arrive at
\begin{equation}
  \phi_j=-Q_j/e^2N_j \mathrm{ .}
  \label{EQ_2}
\end{equation}
Adding for the charge $Q_j$ the values given by (\ref{EQ_1}),
yields
\begin{equation}
  \phi_j=\frac{\epsilon_0(E_{j-1}-E_j)}{e^2N_j}\mathrm{ .}
  \label{EQ_3}
\end{equation}
A self-consistency relation can now be formulated by noting that the difference of the electro\-static potentials
determines the electrical fields, i.e. $\phi_{j}-\phi_{j-1}=E_jd$. This leads to
\begin{equation}
  E_j=\frac{\epsilon}{e^2 Nd}\left\{E_{j+1}-2E_j+E_{j-1}\right\} \mathrm{ ,}
  \label{EQ_4}
\end{equation}
where the density-of-state has been assumed to be constant and given be $N$. This equation
has exponentially decaying solutions of the form
\begin{equation}
  E_j=E_0 e^{-dj/\lambda} \mathrm{ ,}
  \label{EQ_5}
\end{equation}
where $\lambda$ denotes the inter\-layer screening length. Placing (\ref{EQ_5}) into (\ref{EQ_4}) yields
the following condition: $cosh(d/\lambda)-1=e^2Nd/2\epsilon_0$. This equation determines the screening length $\lambda$.
In order to quantify $\lambda$, the areal (2d) density-of-states $N$ is required. We estimate $N$ from the known
3d value $N_0$ of graphite to $N=N_0 d$. With
\mbox{$N_0=5.2\cdot 10^{20}$\,cm$^{-3}$/$e$V} and the inter\-layer distance \mbox{$d=0.34$\,nm} \cite{McClurePR57},
one surprisingly obtains $d/\lambda\cong 1$, i.e. \mbox{$\lambda\cong 0.34$\,nm}.
Eventually, we can place the solutions $E_j$ in (\ref{EQ_5}) back into the equations (\ref{EQ_2}) and (\ref{EQ_3}) to obtain
the charge in layer $j$:
\begin{equation}
  Q_j = -Q_0 e^{-(d/\lambda)(j-1)}\left\{1 - e^{-d/\lambda}\right\} \mathrm{ .}
  \label{EQ_6}
\end{equation}
Obviously, if screening is strong ($\lambda << d$), $|Q_1| \simeq |Q_0|$, whereas $|Q_{j>1}|<<|Q_0|$. In the opposite case
of weak screening ($\lambda >> d$), the charge decays slowly and is given for not too large $j$ by
$|Q_j|\simeq |Q_0|d/\lambda$.
In the present situation with $\lambda \cong d$, $63$\,\% of the charge is in the first layer, $23$\,\% in the second,
and the third layer already carries less than $10$\,\%. Hence, this model of an infinite stack of layers should work well for
few layer graphene stacks when the number of layers $N_L \gtequiv 3$.
The parameter $Q_0$ in (\ref{EQ_6}) can be related to the applied gate voltage $V_g$ using the relation
$\phi_0-\phi_1=V_g-\phi_1= E_0 t$. We then arrive at a relation between the charge $Q_0$ on the gate and the
gate voltage $V_g$:
\begin{equation}
  Q_0 = \frac{V_g}{1/C_Q + 1/C_g} \mathrm{ ,}
  \label{EQ_7}
\end{equation}
where $C_g$ is the geometrical gate capacitance and $C_Q$ is an effective chemical capacitance (sometimes also termed
quantum capacitance), both taken per unit area. $C_Q$ is given by
\begin{equation}
  C_Q = \frac{e^2 N}{1-e^{-d/\lambda}} \mathrm{ .}
  \label{EQ_7b}
\end{equation}
Equation~\ref{EQ_7} shows that this relation represents a series connection of $C_g$ with an effective quantum capacitance
$C_Q$ as depicted in figure~2b. The latter can also be derived from the circuit shown in figure~2c consisting of an infinite series of inter\-layer
capacitances $C_i$ and intra\-layer quantum capacitances $C_q=e^2 N$~\cite{Lorke2006}.
We next estimate the two capacitances $C_g$ and $C_q$.
Taking $\epsilon = 3.9$ for the relative permittivity of SiO$_2$ yields \mbox{$C_g = 1.1 \cdot 10^{-4}$\,F/m$^2$}. With
the aerial density of states $N$, which we estimated before from known graphite values, we obtain \mbox{$C_q = 2.8 \cdot 10^{-2}$\,F/m$^2$}.
Hence, $C_q \gg C_g$ by more than two orders of magnitude. Since the smaller capacitance counts in a series connection
as the one shown in figure~2b, the relation between gate charge $Q_0$ and gate-voltage $V_g$ is to a very good accuracy
given by the normal one $Q_0\cong C_g V_g$.

Having analysed the screening problem, we can calculate the sheet conductivity of the whole stack. We assume to be in the
diffusive limit and use the Einstein equation, which
relates the conductivity $\sigma_j$ to the density-of-state in layer $j$ and the diffusion constant $D$. In a Fermi
gas the diffusion coefficient is given by $D=v_F^2\tau/2$, where $v_F$ is the Fermi velocity, in graphene equal to \mbox{$10^6$\,m/s},
and $\tau$ the scattering time. For simplicity we assume that $D$ is constant. This is an approximation, as it is known that
there are surface effects and it is plausible that adsorbates must have the biggest effect on the mobility of the first layer.
With this assumption we can write for the total conductance $G$ of the FLGNR
\begin{equation}
  G = \frac{W}{L} e^2D \sum_{j=1}^\infty N_j \mathrm{ ,}
  \label{EQ_8}
\end{equation}
where $W$ and $L$ are the width and length of FLGNR, respectively. The gate-dependence of the conductance is due to the
energy dependence in the density-of-states $N_j(E)$ which we have to add now here. In an ideal single graphene layer
the energy dispersion relation is given by $E=\hbar v_F|\vec{k}|$, where $\vec{k}$ denotes the wave\-vector in two dimension.
This dispersion results in a density-of-state $N$ given by~\cite{Zhang05}
\begin{equation}
  N(E) = \frac{2}{\pi}\frac{|E|}{(\hbar v_F)^2} \mathrm{ .}
  \label{EQ_9}
\end{equation}
This density-of-state goes to zero for $E \rightarrow 0$. We have however assumed a finite density-of state $N_0d$ at the
charge neutrality point (CNP, i.e. $E=0$), which is caused by the inter\-layer overlap. To interpolate between the
two regimes, we write:
\begin{equation}
  N(E) = N_0d + \frac{2}{\pi}\frac{|E|}{(\hbar v_F)^2} = N_0d(1+\beta |E|) \mathrm{ .}
  \label{EQ_10}
\end{equation}
The slope $\beta$ is not a free parameter, but determined by the above equation. We obtain
\mbox{$\beta = 8.3$\,$/$eV}.
Since the energy $E$ in (\ref{EQ_9}) denotes the chemical potential, we can replace it with the electro\-static one. This
leads together with (\ref{EQ_8}) to:
\begin{equation}
  G = \frac{W}{L} e^2 D N_0d \sum_{j=1}^\infty (1+\beta|e\phi_j|) \mathrm{ .}
  \label{EQ_11}
\end{equation}
Adding the explicit expressions for $\phi_j (V_g)$, the result is
\begin{equation}
  G = \frac{W}{L}e^2 D N_0d\left\{ N_L + \left(1-e^{-d/\lambda}\right)^{-1} \beta \frac{eV_g}{1+C_q/C_g} \right\}\mathrm{ .}
  \label{EQ_12}
\end{equation}
By dividing with the minimum conductance $G_{min}$ at $V_g=0$, the dependence on the diffusion coefficient drops out. In the practical
limit of $C_q \gg C_g$ the gate voltage change of $G/G_{min}$ is given by the final simple result:
\begin{equation}
  \frac{\partial}{\partial V_g}\left(\frac{G}{G_{min}}\right)=\frac{\beta e}{N_L}\frac{C_g}{C_q}\mathrm{ .}
  \label{EQ_13}
\end{equation}
We can apply this result to figure~2. Equation~\ref{EQ_13} predicts a relative change of \mbox{$3.3$\,\%/V} for $N_L=1$.
We measure a change of \mbox{$1.2$\,\%/V} on the right and \mbox{$0.32$\,\%/V} on the left. These two slopes correspond to
$N_L=3$ on the right and $N_L=10$ on the left. The model clearly shows that we are dealing with a number of layers, but that
we are in the regime of few layers with $N_L \ltequiv 10$. Due to the different slopes for voltages smaller or larger than the CNP,
a more accurate estimate for $N_L$ cannot be given. However, we stress that physically we deal with one FLGNR with one
given $N_L$ number. The reduced change on the gate-voltage to the left of the CNP suggests that the carrier density is
markedly increased on the hole side leading to an enhanced quantum capacitance. As the CNP is strongly on the positive side,
the FLGNR is substantially hole doped. This hole doping may be caused by states that are induced by the source and drain contacts.

The dependence of conductance $G(V_g)$ on the gate voltage $V_g$ can also be used to determine the mobility $\mu$.
Taking the derivative of equation (\ref{EQ_12}) versus $V_g$ and using the relation $C_q \gg C_g$ yields:
\begin{equation}
  \frac{\partial G}{\partial V_g} =\frac{W}{L}De\beta C_g \mathrm{ .}
  \label{EQ_14}
\end{equation}
With the mobility $\mu$ one may also write the conductance as
\begin{equation}
  G=\frac{W}{L}\mu Q\mathrm{ ,}
  \label{EQ_15}
\end{equation}
where $Q$ is the effective carrier density. Since $C_q \gg C_g$, the gate-voltage induced carrier density
is given by $Q\cong C_g V_g = Q_0$. This leads to
\begin{equation}
  \frac{\partial G}{\partial V_g} = \frac{W}{L}\mu C_g \mathrm{ .}
  \label{EQ_16}
\end{equation}
By comparing this equation with equation (\ref{EQ_14}), we deduce the relation $\mu=De\beta$.
Taking the experimental value for $\Delta G/\Delta V_g \simeq 1.2$\,$e^2/h$ per $30$\,V, equation~(\ref{EQ_16}) provides the
value \mbox{$\mu=140$\,cm$^2$/Vs} for the electron mobility. $D$ follows then to be \mbox{$\approx 17$\,cm$^2$/s}, from
which we estimate by virtue of $D=v_F l_e/2$ the scattering mean-free path to $l_e\approx 3.5$\,nm.
With $l_e < L,W$, we conclude that the device is diffusive as anticipated in the beginning.

\section{Conductance measurements}

Upon cooling the sample below the critical temperature of bulk Al, \mbox{$T^{bulk}_{c}=1.2$\,K},
an energy gap \mbox{$\Delta^{bulk}\approx 200$\,$\mu$eV} opens in the density of states of Al.
At the base temperature of \mbox{$230$\,mK} the transport properties of the device are deduced by measuring
the two-terminal differential conductance $G$ as a function of source-drain $V_{sd}$ and back-gate voltage $V_{g}$,
represented by the grey-scale plot in figure~3a. The differential conductance $G$ was obtained by superimposing an
AC excitation voltage of \mbox{$20$\,$\mu$V} onto the DC part of the source-drain voltage and measuring
the pre-amplified AC current with a lock-in amplifier. For each back gate voltage in the range from $-10$\,V to $20$\,V
$V_{sd}$ is swept from $-0.8$\,mV to $0.8$\,mV.

In the grey-scale plot shown in figure~3a we observe a reproducible pattern of conductance fluctuations which is modulated
by both $V_{g}$ and $V_{sd}$. The amplitude of conductance fluctuations clearly diminishes as $V_{sd}$ is increased.
In contrast, as a function of back-gate voltage the fluctuations have a homogeneous amplitude, similar to what is seen in figure~1.
As compared to the data in figure~1, the amplitude is larger in figure~3a due to the lower temperature. This dependence of conductance
fluctuations points to so-called `universal conductance fluctuations' (UCF), which are caused by quantum interference effects~\cite{Lee87}.

\begin{figure}[htb]
  \begin{center}
  \includegraphics[width=1\linewidth]{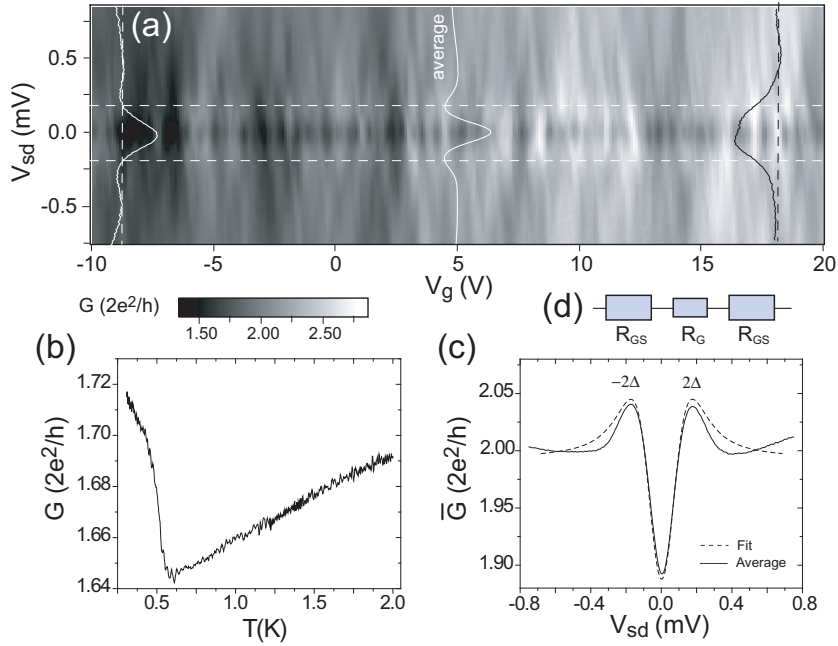}
    \caption{\textbf{a}) Grey-scale plot of the two-terminal conductance $G=dI/dV$ as a function of the back-gate
      voltage $V_{g}$ and the source drain voltage $V_{sd}$ at $T = 230 mK$. The dashed white lines
      indicate the position of twice the super\-conducting gap $2\Delta /e$. An example of a conductance spectra with
      dip around zero bias is shown on the left, whereas one with a peak is shown on the right with the average curve placed in the middle of the plot.
      \textbf{b}) Temperature dependence of the linear conductance taken during the sample cool-down at $V_{g}=0$.
      (\textbf{c}) The conductance averaged over all back-gate curves $\overline{G}(V_{sd})$ as a function of $V_{sd}$.
      (\textbf{d}) In our system the interface resistances between the graphene ad super\-conductor dominate over the internal
      resistance of the FLGNR. This is illustrated in the resistor network.
    }
  \end{center}
\end{figure}

In addition to UCF, the presence of super\-conductivity is evident from the pronounced dark (low) conductance band
around zero bias in between the $2\Delta$-lines, which are marked in figure~3a by white dashed lines. This band disappears
at temperatures above the critical temperature $T_c$, as well as in a magnetic field larger than the critical field.
The band of reduced conductance shows, that the conductance is on average lower at low bias than it is at high bias
(normal state). We illustrate this by overlaying the average conductance $\overline{G}(V_{sd})$ versus
$V_{sd}$ in the middle of the grey-scale plot. This, however, does not mean that all individual $G(V_{sd})$ curves
show dips at zero source-drain voltage. Though the majority of $G(V_{sd})$ curves must display dips, such as
the example overlayed on the left side of the grey-scale plot, there is a significant number of curves that
display peaks around zero bias instead. An example is shown by the right overlayed curve (see also figure~4).

To quantify the energy scale at which the super\-conductivity related effects become important,
we extract the critical temperature $T_c$ in the following. The conductance $G(T)$ as a function of temperature $T$
was measured during the cool down at $V_{g}=0$. This is shown in figure~3b. For this gate-voltage value, $G(T)$ decreases with temperature,
reaching a minimum at about $0.6$\,K. Below this temperature the conductance sharply increased
reaching a value of $1.72 \cdot 2 e^{2}/h$ at the base temperature. We take the cross-over temperature at $0.6$\,K
as the transition temperature $T_c$ to the super\-conducting state. We note that the bulk value for Al is \mbox{$T_c=1.2$\,K}.
The substantial reduction of $T_{c}$ of $0.6K$ in our super\-conductor-graphene-super\-conducutor (S-G-S) device as compared to the bulk value
may be due to the inverse proximity effect from the two normal metal Ti layers with thicknesses of
$5$\,nm and $20$\,nm  surrounding the Al~\cite{Tinkham}.

Figure~3c shows the average conductance $\overline{G}(V_{sd})$ as a function of $V_{sd}$.
This curve is obtained as the mean of all individual conductance curves measured at a fixed $V_{g}$ value,
after a small linear background was subtracted from the grey-scale plot in figure~3a.
In order to fit the data to the BTK model~\cite{BTK}, we need to know how the voltage drops in our S-G-S device.
To estimate this, we evaluate the resistor model shown in figure~3d. The total resistance is divided into two
interface-graphene resistors $R_{GS}$, which on both sides are taken to be equal, and the intrinsic resistance $R_G$
of the FLGNR connected in series. We compare the measured minimum conductance $G_{min}$ with the expected
minimum conductivity $\sigma _{min}$ of a single sheet of graphene. To do so, we will use the number of
graphene layers in our FLGNR which we estimated in the previous section to be between $3$ and $10$.

Unlike the electrons in a two-dimensional electron gas with quadratic dispersion,
the Dirac-electrons do not localize, even not for the smallest possible carrier density.
Instead, there is a minimum conductivity $\sigma_{min}$ which is reached at the CNP.
In experiments with graphene layers of intermediate disorder, $\sigma_{min}$ has been found to
be $\approx 4$\,$e^2/h$~\cite{NovoselovNat05}. In the ballistic regime $\sigma_{min}(4/\pi)$\,$e^2/h$
has been predicted~\cite{Trauzettel2006}. This value, which is reduced by a factor of $\pi$, has not yet been
demonstrated experimentally, even not in the highest quality suspended graphene devices~\cite{NatechAndrei2008}.

The estimate of the elastic mean-free path in the previous chapter has shown that we are in the diffusive limit.
In our experiment $G_{min}\sim 3.2$\,$e^2/h$, which relates to $\ltequiv 1$\,$e^2/h$ per layer. Since this value is appreciably
smaller (by $\approx$ a factor of $4$) than $\sigma_{min}$ in the diffusive limit~~\cite{NovoselovNat05}, the total resistance
must be dominated by the interface resistances, i.e. $2R_{GS} > R_G$. Taking this into account, we
simplify our network by dropping $R_G$ altogether. We therefore assume that the applied source-drain voltage $V_{sd}$ drops
symmetrically over the source and drain contact. The voltage drop $V_{GS}$ across one interface
is then half of the voltage $V_{sd}$ applied across the junction and, hence, $G_{GS}=2G$.

We are now in the position to model the measured average conductance curve $\overline{G}(V_{sd})$ in figure~3c
using the BTK model~\cite{Dynes78, BTK}. This fit, which is shown in figure~3c by the dashed curve,
provides as one parameter the  super\-conducting energy gap \mbox{$\Delta^{av} = 60$\,$\mu$eV}.
Other fitting parameters deduced from the model are the effective barrier strength $Z\approx 0.6$, and an
inelastic broadening parameter \mbox{$\Gamma \approx 10$\,$\mu$eV}.
The broadening parameter is introduced as an imaginary energy term in the BCS density of states,
$\rho (E, \Gamma)=(E-i \Gamma)/[(E-i \Gamma)- \Delta ^{2}]^{1/2}$ with $E$ being the energy of
quasi\-particles measured from the Fermi energy~\cite{Dynes78}. The super\-conducting energy gap $\Delta^{av}$ can
be compared to the previously deduced transition temperature $T_c=0.6$\,K. Taking the BCS relation
$\Delta =1.76k_{B}T_{c}$~\cite{Tinkham}, one obtains \mbox{$\Delta\approx 90$\,$\mu$eV}. Good agreement
between the two numbers is found. The consistent numbers support our simplified approximation that most of the
voltage drops over the superconductor-graphene contacts.

\begin{figure}[htb]
  \begin{center}
  \includegraphics[width=0.5\linewidth]{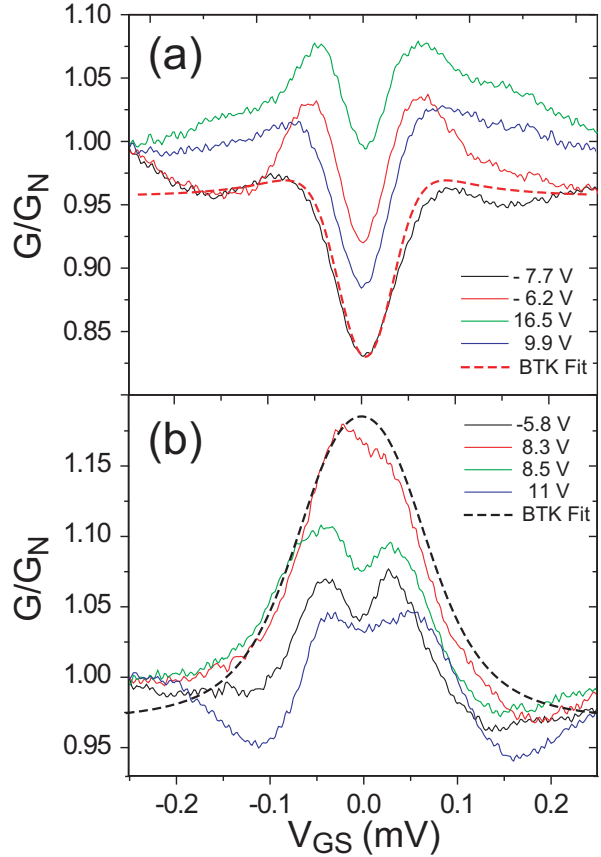}
    \caption{
    Typical differential conductance curves $G(V_{GS})$ for different constant back-gate voltages $V_{g}$
    displayed as a function of the voltage $V_{GS}$ that drops over a single graphene-super\-conductor junction.
    $V_{GS}\simeq V_{sd}/2$.
    (\textbf{a}) In the bias window $|V_{GS}|< \Delta /e$ a dip is observed in $G(V_{GS})$ for $V_g = -7.7$, $-6.2$, $16.5$, and $9.9$\,V,
    whereas ({\bf b}) a peak appears for $V_{g}= -5.8$, $8.3$, $8.5$,and $11$\,V.
    All curves are normalized to the large bias conductance values $G_N$ which are
    $1.76$ ($-7.7$), $1.85$ ($-6.2$), $2.44$ ($16.5$), $2.38$ ($9.9$), $1.92$ ($-5.8$),
    $2.23$ ($8.3$), $2.25$ ($8.5$), $2.48$ ($11$) in units of $2e^2/h$, where the number in bracket denotes $V_g$.
    }
  \end{center}
\end{figure}

Figure~4 provides examples of individual conductance curves $G(V_{GS})$ taken for different values of $V_g$.
Note, $V_{GS}$ is the voltage over one graphene-super\-conducutor junction and taken to be half of $V_{sd}$.
As we have emphasised before, $G(V_{GS})$ is suppressed on average in the sub-gap region for small source-drain voltages
$V_{sd}$ because the majority of curves show dips around zero $V_{sd}$. Examples of such curves are shown in
figure~4a for back-gate values $V_{g}= -7.7$\,V, $-6.2$\,V, $16.5$\,V and $9.9$\,V.
However, a substantial number of curves show peaks. Examples are given in figure~4b taken at
$V_{g}= -5.8$\,V, $8.3$\,V, $8.5$\,V and $11$\,V. This shows that Andreev reflection can dominate sub-gap transport
at the G-S interface and thereby enhance $G$.
After fitting individual curves to the BTK model~\cite{Dynes78, BTK}  (dashed lines in (a) and (b)),
we extract a super\-conducting gap of \mbox{$\Delta/e\approx 65$\,$\mu$V}, which agrees very well with
\mbox{$\Delta^{av} = 60$\,$\mu$eV} deduced for the average curve before.
The inelastic broadening parameter $\Gamma$ ranges between \mbox{$10-20$\,$\mu$V} and for the barrier strength $Z$
we obtain $\approx 1$ for curves with dips and $Z \approx 0.3$ for traces with peaks.
This agrees with the notion that the conductance of a super\-conductor-normal interface
can be enhanced in the super\-conducting state, provided the interface transmission is sufficiently high, and hence, the
barrier strength $Z$ is sufficiently small.

We now turn our attention to the bias-dependence of the conductance fluctuations.
From the conductance data in figure~3 obtained at \mbox{$230$\,mK} we determine the rms conductance variance
for a fixed source-drain voltage $V_{sd}$ by a statistical average over all $G(V_g)$ values in the gate-voltage range.
We assume that the large gate-voltage window provides a sufficiently large ensemble of different disorder configurations.
In order to determine the root-mean square variance $\delta G$ a linear background has been subtracted from
$G(V_{g})$ leaving us with values $\Delta G(V_{g})$ with a zero mean. The rms variance $\delta G$
of the whole two-terminal device is then given by $\langle \Delta G ^2 \rangle^{1/2}$. For comparison with
theory it can be convenient to relate the conductance variance to one graphene-super\-conductor (G-S) interface, which we
denote by $\delta G_{GS}$. If we assume that the two G-S contacts fluctuate independently $\delta G_{GS} = \sqrt{2}(\delta G)$.
This is plotted in figure~5 as a function of voltage $V_{GS}=V_{sd}/2$ dropping over one contact.

\begin{figure}[htb]
  \begin{center}
  \includegraphics[width=0.6\linewidth]{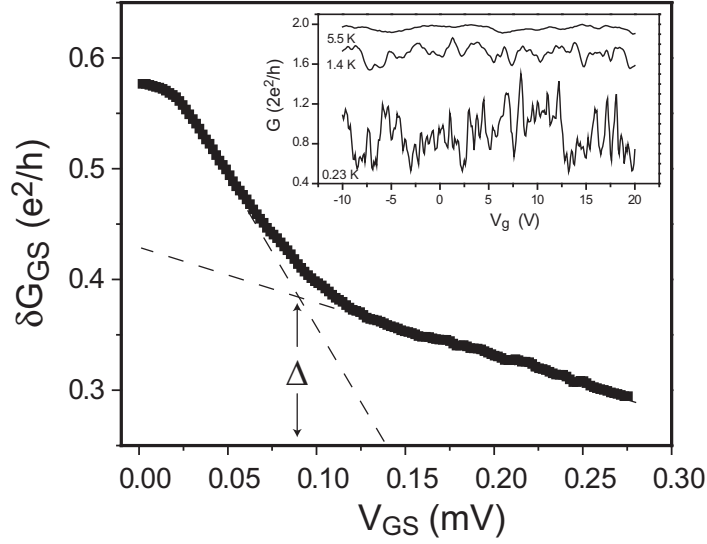}
    \caption{
    The conductance variance of a single graphene-super\-conductor (G-S) interface $\delta G_{GS}$
    (root mean square of the deviation of $G(V_{sd})$ relative to linear background times $\sqrt{2}$)
    as a function of the voltage drop across the G-S interface.
    $\delta G (V_{GS})$ displays a pronounced increase below a cross-over voltage that agrees with the
    super\-conducting gap value $\Delta$.
    The inset displays $G(V_g)$ measured at different temperatures $5.5$\,K, $1.4$\,K and $0.23$\,K,
    after subtracting a linear back\-ground for comparison.
    The $5.5$\,K curve is shifted by $0.2\cdot 2e^2/h$ and the $0.23$\,K one by $-1.2\cdot 2e^2/h$ for clarity.
    }
  \end{center}
\end{figure}

In figure~5 we note two distinct regions with a clear cross-over at the energy \mbox{$eV_{GS}\sim 90$\,$\mu$eV} (arrow),
which coincides with the previously estimated super\-conducting energy gap $\Delta$ of the Ti/Al/Ti tri-layer.
In both regions $\delta G_{GS}$ decreases with the applied bias. However, the decreasing rates of $\delta G_{GS}$
are different for the two regions. In the high bias region, $V_{GS} > \Delta /e$, $\delta G_{GS}$ decreases
four times slower than in the region below the super\-conducting energy gap $\Delta$.
In the latter region, $\delta G_{GS}$ starts saturating as the voltage drop across the junction $V_{GS}$
approaches \mbox{$k_{B}T/e \approx 20$\,$\mu$V} and takes a finite value of $\delta G_{GS,230 mK}=0.58 e^{2}/h$
at zero bias. In addition, from the inset where conductance traces at $5.5$\,K, $1.4$\,K and $230$\,mK are shown,
we find for the conductance variance at $1.4\,K$ the value $\delta G_{GN,1.4\,K}=0.19$\,$e^{2}/h$ and at $5.5\,K$
$\delta G_{GN,5.5\,K}=0.06$\,$e^{2}/h$, where the index $N$ denotes that Al is in the normal state.

It is clear that the observed increase in $\delta G_{GS}$ is due to the presence of the super\-conductor and
Andreev reflections at the G-S interface. At zero temperature and zero magnetic field theory predicts an increase of $2.07$
for the conductance variance of a single super\-conducting-normal metal junction in the super\-conducting
relative to the normal state~\cite{BeenakkerNato}.
A simple estimate provides from our data an enhancement of $\approx 1.4$, which is smaller than the maximum that can be expected.
This estimate is obtained by linearly extrapolating $\delta G$ from the high bias down to zero bias
(lower dashed line in figure~5). This extrapolation yields $\delta G = 0.3$\,$e^{2}/h$ for the variance $\delta G$ of the whole
two-terminal sample.

The reduced magnitude of the conductance variance ratio $\delta G_{GS}/\delta G_{GN}$ is not unexpected, taking the energy of the
lowest temperature and the finite AC-bias modulation into account, which both lead to saturation of $\delta G_{GS}$
at the lowest bias voltage in figure~5. The reduction of the saturation value $\delta G$ from the expected
`universal' value can be caused by finite coherence. In the context of interference correction two parameters are important
at zero magnetic field: the thermal length $l_T$ and the phase-breaking length $l_{\varphi}$. The thermal length, sometimes also
termed the coherence length, is given by $l_{T}=(\hbar D/k_{B}T)^{1/2}$. With $D= 17$\,cm$^2$/s, we obtain $l_T\approx 240$\,nm.
At the lowest temperature, the devices crosses therefore over into the fully coherent regime. Hence, the reduced value
of $\delta G$ must primarily originate from a finite phase-breaking length $l_{\varphi} < L$.

To compare further with theory, a value of $\delta G = 0.69\sqrt{W/L}$\,$e^2/h$ was predicted for a graphene ribbon with
normal metal contacts~\cite{RycerzEPL2007}. For the geometry of our device this would translate into
$\delta G \approx 0.6$\,$e^2/h$. Our estimated normal state $\delta G$ value is two times smaller.
This can be used to estimate an effective coherence length $l_{\varphi}$ by using standard averaging along the
length of the ribbon~\cite{BeenakkerVHauten}, yielding $G \sim 0.69\sqrt{W/L}(L_{\varphi}/L)^{3/2}\,e^2/h$.
With $L$ being \mbox{$225$\,nm}, we estimate a phase-breaking length of \mbox{$L_{\varphi}\approx 80$\,nm} at \mbox{$230$\,mK}
and consequently a phase-breaking time \mbox{$\tau_{\varphi}\approx 4$\,ps}.

The bias dependence of $\delta G$ has not been studied systematically in graphene-super\-conductor devices. However
some data are available in the literature. In the recent experiments on single graphene layer contacted with Pt/Ta
super\-conducting leads the conductance variance at $60$\,mK was found to be $\delta G_{GS}=2.4$\,$e^2/h$~\cite{OjedaPRB2009}.
Though this value is substantially larger than ours, a comparison has to take the larger width $W$ into account
($W=2.7$\,$\mu$m and $L=330$\,nm). It turns out that also in this experiment the measured value is lower than the full coherence value,
which we estimate to $4.3$\,$e^2/h$ using the expression $2.07\times 0.69 \sqrt{W/L} e^2/h$.

In nano\-structures made of InAs nano\-wires contacted with Ti/Al leads~\cite{DohNL08,DohKorean09,JespersenArxive09}
the saturation value of the conductance variance in the super\-conducting state was found to be
$\sim 0.8$\,$e^2/h$ at $22$\,mK \cite{DohNL08, DohKorean09} and $\sim 0.47$\,$e^{2}/h$ in the normal state
yielding an enhancement of $\sim 1.6$, whereas in \cite{JespersenArxive09} at \mbox{$300$\,mK}
$\delta G \sim 0.7$\,$e^2/h$ in the super\-conducting state was found with an enhancement factor of $\sim 1.5$ compared to the
normal state. Both the normal state values and the enhancement factors for the super\-conducting state of these
results are in good agreement with our observations in a Ti/Al contacted few layer graphene nano\-ribbon.

In recent experiments with Niobium contacted InAs nano\-wires a surprisingly large enhancement factor of $\sim 30$ was
found~\cite{JespersenArxive09}. The reason for such a strong enhancement is that the coherence length of the nano\-wire
and the dephasing length $l_{\varphi}$ is much larger than the distance between the contacts, so that
multiple Andreev charge transfer which contributes to fluctuations can occur at the interface.
The increased conductance variance signals the transition into the super\-conducting state of the whole device.

\section{Summary and discussions}

In conclusion, our measurements of a few-layer graphene nano\-ribbon contacted with Ti/Al leads show pronounced
UCF-type conductance fluctuations. We observe a decrease of the conductance variance with applied source-drain
voltage with a characteristic cross over for bias voltages corresponding to the super\-conducting energy gap $\Delta$.
For voltages below $\Delta/e$ the conductance variance is enhanced by a factor ranging between $1.4$ and $1.8$ which is
close to the theoretically predicted value of $2.07$. The finite phase-breaking length $L_{\varphi}< L$
at the base measurement temperature of $230$\,mK is the reason for the remaining discrepancy.

\ack

This work is financially supported by the NCCR on Nanoscale Science, the Swiss-NSF and EU-FP6-IST project HYSWITCH.
We are grateful to the PSI (Paul Scherrer Institute) for the thermal oxidation of our Si wafers.


\section*{References}


\begin{thebibliography}{99}
\bibitem{NovoselovSc04}
  Novoselov K S, Geim A K, Morozov S V, Jiang D, Zhang Y, Dubonos S V, Grigorieva I V and Firsov A A 2004 \textit{Science} {\bf 306} 666
\bibitem{TheRise}
  Geim A K and Novoselov K S 2007 \textit{Nature Materials} {\bf 6} 183
\bibitem{Gprospects}
  Geim A K 2009 \textit{Science} {\bf 324} 1530
\bibitem{Hee07}
  Heersche  H B, Jarillo-Herrero P, Oostinga J B, Vandersypen L M K and Morpurgo A F 2007 \textit{Nature} {\bf 446} 56
%
%
\bibitem{XuDu2008}
Du Xu, Skachko Ivan and Andrei Eva Y 2008 \textit{Phys. Rev.} B {\bf 77} 184507
\bibitem{MorozovPRL06}
  Morozov S V, Novoselov K S, Katsnelson M I, Schedin F, Ponomarenko L A, Jiang D and Geim A K 2006 \textit{Phys. Rev. Lett.} {\bf 97} 016801
\bibitem{HBHeerscheEPJSpec2007}
  Heersche H B, Jarillo-Herrero P, Oostinga J B, Vandersypen L M K and Morpurgo A F 2007 \textit{Eur. Phys. J. Special Topics } {\bf 148} 27-37
\bibitem{OjedaPRB2009}
  Ojeda-Aristizabal C, Ferrier M, Gu{\'e}ron S and Bouchiat H 2009 \textit{Phys. Rev.} B {\bf 79} 165436
\bibitem{Lee87}
  Lee P A, Douglas Stone A, Fukuyama H 1987  \textit{Phys. Rev.} B {\bf 35} 1039
\bibitem{RycerzEPL2007}
  Rycerz A, Tworzydlo J and Beenakker C W J 2007 \textit{Europhys. Lett.} {\bf 79} 57003
\bibitem{FuhrerNNanTech08}
  Chen Sion-Hao and Fuhrer Michael S 2008 \textit{Nature Nanotechnology} {\bf 3} 206
\bibitem{PBlakeContacts}
  Blake P, Yang R, Morozov S V, Schedin F, Ponomarenko L A, Zhukov A A, Grigorieva I V, Novoselov K S, Geim A K 2009
  \textit{Solid State Commun.} {\bf 149} 1068-1071
\bibitem{contactLeeNNT08}
  Lee Eduardo J H, Balasubramanian Kannan, Weitz Ralf Thomas, Burghard Marko and Kern Klaus 2008 \textit{Nature Nanotechnology } {\bf 3} 486 - 490
\bibitem{GuineaPRB08}
  Guinea F 2008 Phys. Rev. B {\bf 77} 205421
\bibitem{BeenakkerNato}
   Beenakker C W J 1995 \emph {Mesoscopic Quantum Physics} (Amsterdam: North-Holland)
\bibitem{Zhang05}
  Zhang Yuanbo, Small Joshua P, Pontius William V and Kim Phillip 2005 \textit{Appl. Phys. Lett.} {\bf 86} 073104
\bibitem{TIhn2008}
  Ihn T, Graf D, Molitor F, Stampfer C, Ensslin K 2008 \textit{Physica} E {\bf 40} 1851-1854
\bibitem{McClurePR57}
  J. W. McClure 1957 \textit{Phys. Rev.} {\bf 108} 612
\bibitem{Lorke2006}
  Russ M, Meier C, Lorke A, Reuter D and Wieck A D 2006 \textit{Phys. Rev.} B {\bf 73} 115334
\bibitem{Tinkham}
  Michael Tinkham 2004 \emph{Introduction to superconductivity}, (New York: Dover Publications Inc.)
\bibitem{BTK}
  Blonder G E, Tinkham M and Klapwijk T M 1982 \textit{Phys. Rev.} B {\bf 25} 4515
\bibitem{NovoselovNat05}
  Novoselov K S, Geim A K, Morozov S V, Jiang D, Katsnelson M I, Grigorieva I V, Dubonos S V and Firsov A A 2005 \textit{Nature} {\bf 438} 197-200
\bibitem{Trauzettel2006}
  Tworzydlo J, Trauzettel B, Titov M, Rycerz A and Beenakker C W J 2006 \textit{Phys. Rev. Lett.} {\bf 96} 246802
\bibitem{NatechAndrei2008}
Du Xu, Skachko Ivan, Barker Anthony and Andrei Eva Y 2008 \textit{Nature Nanotechnology} {\bf 3} 491
\bibitem{Dynes78}
  Dynes R C, Narayanamurti V and Garno J P 1978 \textit{Phys. Rev. Lett.} {\bf 41} 1509 - 1512
\bibitem{BeenakkerVHauten}
  Beenakker C W J and van Houten H 1991 \textit{Solid State Physics} {\bf 44} 1-228
\bibitem{DohNL08}
  Doh Yong-Joo, De Franceschi Silvano, Bakkers Erik P A M, Kouwenhoven Leo P 2008 \textit{Nano Letters }{\bf 8} 4098-4102
\bibitem{DohKorean09}
  Doh Yong-Joo, Roest Aarnoud L, Bakkers Erik P A M, De Franceschi Silvano and Kouwenhoven Leo P 2009
  \textit{Journal of the Korean Physical Society} {\bf 54} 135-139
 \bibitem{JespersenArxive09}
 Jespersen T S, Polianski M L, Soerensen C B, Flensberg K, Nygaard J 2009 arXiv:0901.4242

\end{thebibliography}
\end{document}